# The Workshop — Implementing Well Structured Enterprise Applications
## by the Example of Implementing the Key Authority

A. Wiesmaier, V. Karatsiolis, M. Lippert, J. Buchmann
Technische Universität Darmstadt, Department of Computer Science
Hochschulstr. 10, 64289 Darmstadt, Germany
[wiesmaie|karatsio|mal|buchmann]@cdc.informatik.tu–darmstadt.de

*Abstract*— We specify an abstraction layer to be used between an enterprise application and the utilized enterprise framework (like J2EE or .NET). This specification is called the Workshop. It provides an intuitive metaphor supporting the programmer in designing easy understandable code. We present an implementation of this specification. It is based upon the J2EE framework and is called the JWorkshop. As a proof of concept we implement a special certification authority called the Key Authority based upon the JWorkshop. The mentioned certification authority runs very successfully in a variety of different real world projects.

## I. INTRODUCTION

The Key Authority (KA) is a special version of a certification authority (CA). Among its tasks are signing certificates and signing revocation lists. It is designed especially to enable easy enforcement of secure key management in hierarchical public key infrastructures (PKI). Due to this it has to provide some more functionality than a usual CA. A detailed abstract description of the KA and its functional demands is given in [1]. In opposite to that, the paper on hand only deals with the implementation and the non functional demands of the KA.

A KA software which is sold to many different customers must be able to fulfill the various customer demands. This means it must fit into and be usable within the particular customer environment while fulfilling the customer demands. Such KA software must be flexible in various ways:

Embedabilty: The KA must be adaptable to the existing customer workflow and environment (and not vice versa).

Efficiency: The KA installation must be efficient regarding to the customers use case. It must be able to do its work with an adequate amount of resources.

Scalability: The KA installation must be able to scale with the current load situation. If the amount of work rises it must be possible to increase the power of the installation.

Compatibility: The PKI software must create its products regarding to the customer wishes. This includes the utilized cryptographic primitives, the compatibility to standards or existing proprietary customer formats.

Functionality: The PKI software must offer exactly that functionality which is demanded by the customer.

Security: The desired security level differs with the different purposes of the various trust centers. It must be possible to protect the system with an adequate level of security.

We decided to implement the KA as an enterprise application. Programming enterprise applications is a substantial part of today's software development. This is supported by (usually object oriented) technologies like .NET [2] or J2EE [3]. Those frameworks provide a great help in developing software with certain non functional demands. Those frameworks are quite general. Usually enterprise applications are very big. It is a hard and recurring task to develop a software architecture which solves the assigned tasks with an easy understandable code.

We have chosen the J2EE technology as basis for the KA implementation. The need for an abstraction layer between the KA implementation and the J2EE framework arose soon. This gave the birth to the idea for the Workshop specification. This paper reports about the Workshop specification, its J2EE implementation called the JWorkshop and the implementation of the KA upon it.

The next Section II shows which non functional demands on the KA exist and explains their meaning in the context of KAs. In Section III we explain how we designed the KA to fulfill the mentioned non functional demands. Thereby we introduce the Workshop specification and the JWorkshop implementation. After that, in Section IV, we audit how the mentioned demands are fulfilled by revealing some implementation details. Section V gives some real life examples on existing KA installations as a proof of concept. Finally, in Section VI, we conclude the paper.

## II. NON FUNCTIONAL DEMANDS

The topic of non functional demands is discussed since a couple of years. In most cases they appear in the context of so called enterprise applications. Those demands are usually referred as "the ilities". There are a couple of lists identifying various ilities. Examples derive from the Reference Model Extension Green Paper [4], the OMG–DARPA Workshop on Compositional Software Architectures [5] or Filman's Paper Achieving Ilities [6].

Each list covers a different set of ilities. Some lists are very short while others are very extensive. The definitions of the individual ilities differ from list to list. Sometimes different names are used for the same ility. In some cases an ility is defined (or resolved) by a set of other ilities. Some of the defined ilities are self-evident, like functionality.

Clearly, not all thinkable non functional demands regard the KA. We are concerned with implementing and installing KAs since a couple of years in many different projects. Thereby we learned which demands are laid on a KA. Clearly those demands differ between the various carriers and environments. Below we list ilities for the KA that we have encountered. Thereby we explain which meaning the individual ilities have in the context of KAs. We do not claim that this is an exhaustive list. We give an overview of the most important ilities. Additionally, the criticality of the various ilities differs from project to project depending on the customer wishes and the purpose of the particular KA installation. Due to this the criticality of ilities is not discussed here.

## A. Explicit Ilities

A major part of the ilities where requested directly by the customers. Others derived directly from the purpose of the trust center respective from the environment where the KA was installed.

*1) Availability:* The KA is the instance which signs the issuer products. Thus, it should be available if a certification has to be done. But the PKI will run onward correctly even if the certification service might be unavailable for the moment. The effect will be that no new users can join the PKI for moment. But the KA must be high available for issuing revocations. It must be possible to revoke certificates at any time. Otherwise the correct function of the PKI is not given.

*2) Configurability:* It must be possible to configure the KA instance to the actual desires. This includes static settings which are done usually only at setup time. Those are the URL of the database, the logging targets, the ports where additional hardware is installed and so on. But this also includes dynamic settings which may be regularly changed after setup time and sometimes even at runtime. Examples among other things are the log level and the applied cryptographic primitives (see durability why). It depends on the security level and the customer wishes which parameters shall remain (static or dynamic) configurable after the initial setup.

*3) Durability:* It must be guaranteed that it is possible to run the KA reasonably for a long time. The security of the PKI depends largely upon the security of the underlying cryptographic primitives. The research on cryptanalysis goes on and the power of the computer systems increases. To reflect this, e.g. the German Federal IT Security Agency suggests adequate cryptographic algorithms and parameters regularly. The KA must be able to follow those or similar recommendations and therefore must support the usage of any current and future cryptographic hardware, algorithm and parameter. Even the implementations of the algorithms must be exchangeable, as those could evolve as being weak, too. Durability here is the ability to adapt to the foreseeable weakening of cryptographic primitives by staying up–to–date in cryptographic progress.

*4) Failover:* Durability avoids security crashes by adapting the system to ongoing security issues. Failover in contrast means to be able to keep the system secure even if unforeseen events occur. An example for this occurred recently. The widespread hash algorithm SHA1 evolved as being weaker than assumed [7]. The KA has to offer a fail safe concept which avoids a security crash if an algorithm, the parameters, the implementation or an individual key evolves as being insecure while in use. Failover here is the ability to survive a sudden break of cryptographic primitives.

*5) Interoperability:* The KA has to be able to conform to standards. On the one hand this means the KA must be able to conform to the given international standards. This ensures the interoperability within an international context. But on the other hand this means the KA must be able to conform to proprietary formats which are already used in the target environment. Even if this means to contravene international standards. This ensures the interoperability with existing special customer software.

*6) Manageability:* In order to preserve the ease of use the KA must be manageable from a centralized point. This means starting, stopping, setting parameters or just inspecting the state of the software can be done form a dedicated single user interface. A centralized management also avoids errors with unequal parameter values in different modules of the KA.

*7) Modularity:* In some cases (we had such cases in real projects) the various tasks of the KA have to be executed at completely different places. For example imagine a company running a TC. For sufficient protection of the issuer private keys and high availability of certification and revocation services the major part of the TC is hosted at the computer center. But due to organizational reasons it is better to personalize the hard tokens at the respective staff departments. Thus the KA must be able to be partitioned into autonomous but cooperating sub applications.

*8) Performance:* It must be guaranteed that the system responds in the given time intervals. This demand must be met with an adequate amount of hardware. Some installations are low capacity systems which have to issue only a few certificates a year. An example for this is a national root CA. Others are high capacity systems which have to issue thousands of certificates a day and answer to millions of OCSP [8] requests. Web mail providers might run such systems. Surely it is not economic to setup a high capacity KA for low capacity purposes. And clearly it is not a good idea to setup a low capacity KA for high capacity purposes.

*9) Reliability:* The functionality and the security of the system must be guaranteed. In some cases they even have to be proven. Examples for this are systems for national root CAs. The European directive on digital signatures demands such systems to be highly evaluated.

*10) Retracability:* The KA is an application which is (usually) used for high secure purposes. Thus it must be possible to retrace all actions. E.g. it must be discoverable who the doer of an action was or when an action did take place.

*11) Robustness:* The KA must cope with incorrect applications, crashing systems and other error situations which might occur. Either the KA must be able to repair such error situations or go to a safe state if the correct functionality can not be further guaranteed. This is even more important when the KA is physically shielded from the outer world and has to run a long time behind closed doors. See the TrustSuite project [9] for an example of this.

*12) Scalability:* An existing KA installation may have to deal with different load demands. Imagine a company. In normal case the KA has to produce certificates for new employees or recertify expired keys. In addition some revocations have to be done. Let's say this is the normal load level. But what when a new department is to be integrated into the PKI? Or if the underlying cryptographic algorithms or parameters evolve as weak? In these cases thousands of keys and certificates have to be produced within a short period. This means the load is many times over the normal level. It is not economic to run a high capacity KA the whole time and have the load on a very low level almost all the time. Thus the KA should run in an adequate low capacity mode normally. But if it is necessary to run the KA in a high capacity mode this must be possible.

*13) Security:* For the KA security means that the access policy is enforced. It must not be possible to gain unauthorized access to the system. Clearly, this means that only authorized personnel is allowed to operate the KA. But this also means that the communication within the KA modules and the other trust center components have to be shielded against unauthorized access. And it means that the KA software and the configuration files have to be shielded against unauthorized access. See [1] for more details on the KAs security.

## B. Implicit Ilities

The non functional demands mentioned in this Section do not derive directly from customer requests or purposes of the installed TCs. They derive from the fact that different carriers have different demands. Thus, the end users don't care about them but they are very precious for the developers.

*1) Adaptability:* The functional and non functional demands differ from project to project. Each customer has his own set of demands and particularities. Over the time a lot of implemented functionalities and ilities will be accumulated. The KA software must be designed in a way that it is easy to enable those of them which are needed in the current project, and disable the other ones. Examples are the utilized cryptographic primitives (as RSA or ECDSA), the supported standards (as X.509 or ISIS–MTT), the performance (as high or low capacity), the produced tokens (as chip cards or PKCS#12 files) and more.

*2) Embedabilty:* The KA is a part of a cooperation of PKI modules. It receives its applications from somewhere, sends its products elsewhere and is operated by someone. The TC is probably installed in an existing or predefined organizational environment. Thus, the KA must be able to be embedded in and deal with arbitrary workflows. This means for example the KA must be able to interact with different kinds of data import, data export and user interaction.

*3) Extensibility:* The field of PKI is not static. New standards appear or new cryptographic hardware is developed. A variety of new mechanisms can be expected for the future. And with this a lot of new functional demands from the customers. It is similar with the ilities. New projects might bring new non functional demands with them. It must be easy to extend the KA by functional and non functional demands. In favor without changing the existing code.

*4) Maintainability:* The software has to be maintained. It must be able to fix existing bugs, or adapt the system to new circumstances. It is important that the code is understandable. Not only for the actual programmer but for all programmers which might have to change the code.

*5) Platform independence:* Another demand is the ability to run on different platforms. On the one hand this means the KA must be able to run on various operating systems, with various data bases or application servers. We experienced that the customers mostly want the trust center to run on the systems they are already familiar with. On the other hand this means the KA must be able to deal with various cryptographic hardware types. The customers have different claims regarding e.g. the chip cards or cryptographic cards to be utilized. This depends on the desired security level and the field of application.

## III. DESIGN

A KA which is able to meet the various demands (as required by the respective customer) must be designed very flexible. See Section I for details on the flexibility. This Section explains the design basics of the KA implementation.

### A. J2EE

The KA is programmed in pure Java [10] and utilizes features from J2EE technology, mainly the Enterprise Java Beans (EJB) [11]. While rendering the design for the KA we realized that building the software directly upon the EJB framework would have two disadvantages.

Firstly, the EJB specification is far too general. It leaves the programmer too much space for implementing things. There is nothing like the standard way to implement a component and then plug it into the existing application to cooperate with the existing components. Having many programmers at the project would lead in having many different interfaces and particularities.

Secondly, the EJB technology lacks an intuitive metaphor. It does not support a program structure which can be easily surveyed. Dealing with the raw Java Beans is uncomfortable. They coexist more or less unorganized side by side and it is not easy to see on the first view which of them cooperate and how they do this.

Thus we decided to specify an abstraction layer which solves the mentioned problems. This specification is explained in the next Subsection.

### B. The Workshop

The Workshop is a general specification for an abstraction layer between an enterprise application and the underlying enterprise framework. We kept the specification free from J2EE specific attributes in order to be able to also use it with other enterprise frameworks.

The basic idea is to allow a task to distribute itself instead of forcing its distribution from outside. Each task is implemented as a special class, and is processed by the object flow principle. This means the environment offers suitable workplaces and the task object visits them one by one to fulfill its task. Applications built upon the Workshop will be scalable, fault tolerant and dynamically extendable.

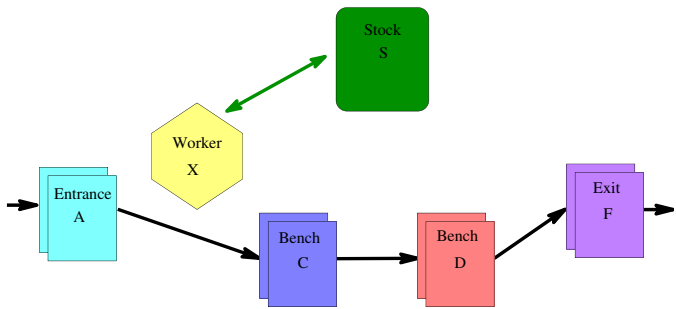

Fig. 1

THE WORKSHOP ARCHITECTURE

Figure 1 shows an overview scheme of the Workshop. An application programmed upon the Workshop follows the metaphor of a worker in a workshop known from the real world. A `Worker` enters the `Workshop` through an `Entrance`. The `Worker` wants to process a work piece that he has brought with him or which he has fetched from the `Stock`. He roams to several `Benches` to perform his job. The `Worker` knows exactly which sub-operations are to be done in which order and on which `Bench`. He is also responsible for a correct exception handling and knows what to do in case of an error. If he needs additional material he fetches it from the `Stock`. Salvageable items are stored in the `Stock`. Finally he leaves through an `Exit`.

To scale the Workshop one can add new instances of the existing `Benches` (on additional hardware) to discharge the existing ones. Or remove redundant instances (and hardware) if the load is low. Adding redundant `Benches` on additional hardware also makes the system tolerating a drop out of a node. This workflow also enables the dynamic extension of the system, as it is possible to add entirely new `Workers` and `Benches`.

A `Worker` knows exactly what task it has to fulfill, how this is done and which `Benches` it needs for that. Each `Worker` instance bears the sole responsibility for the correctness of its process, including a correct exception handling. In order to adjoin a brand new task to the system (that means extending the system) a new `Worker` type is implemented which performs the new job. Sometimes it may then be necessary to implement a new `Bench` (this will be explained in the following paragraph). The existing `Worker` types can persist absolutely untouched. If multiple `Workers` are in the system at the same time, they can work in parallel (if they find free `Benches`). One `Worker` can use at most one `Bench` at a time.

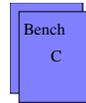 `Benches` act as the necessary work places where the `Workers` do their work. Thereto the `Benches` provide a suitable environment and proper tools for the `Workers`. There must be a convenient `Bench` for each operation a `Worker` has to do. If the system is extended by a new `Worker` type it may be necessary to implement a new `Bench` type. The `Bench` types which are already available are not affected by the new type. There must be at least one instance of each `Bench` type needed by the `Workers` to provide all necessary work places. One Bench can be used at most by one Worker at a time. Multiple `Benches` available at the same time enable multiple `Workers` to work in parallel.

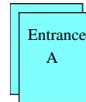 `Entrances` provide the way to bring `Workers` into the Workshop. By conception they stand between the outside and the inside of the system. The `Entrance` receives the necessary data from the outside of the system, checks and interprets it and instantiates an appropriate `Worker`. Thus, a `Worker`'s initial station is always an `Entrance`; from there the `Worker` starts its journey through the system. If a new kind of import is required a new `Entrance` type has to be implemented. The preexisting `Entrances` are not influenced by this. It is possible to have multiple `Entrances` (of the same type or of different types) in the system at the same time. A system must have at least one `Entrance` running at a time to be able to import `Workers`.

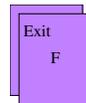 Similar to the `Entrances` the `Exits` live on the border of the inner to the outer of the system. They enable the `Workers` to leave the system when they have finished their work. Leaving here means for a `Worker` to be destructed and converted into a proper export data format. For adding a new form of export one has just to implement a new type of `Exit`. The remaining `Exits` are not affected by this. It is possible to have multiple `Exits` (of the same type or of different types) in the system at the same time. It is necessary to have at least one `Exit` in the system to enable the `Workers` to leave the system.

The `Stock` is a central point in the Workshop. It provides a possibility for the `Workers` to fetch materials from or respectively leave things in the system. There should be always one `Stock` in a system to provide the `Workers` a means to request, fetch, leave or exchange data. A system might have different types of `Stocks` for storing different materials. Having multiple instances of the same `Stock` in a system requires some kind of synchronization between them but enables fault tolerance.

### C. The JWorkshop

The JWorkshop (JWS) is an implementation of the Workshop specification based on the J2EE technology. The

JWS is a general framework and is kept free from KA specific attributes in order to be able to implement arbitrary applications upon it.

To implement the JWorkshop we implemented Enterprise JavaBeans for all Workshop entities (Workers, Benches, ...). When implementing a concrete application one has just to add the desired functionality to the respective classes.

Workers and Stocks are implemented as Entity–Beans. It is possible to have them persistent in the database. Benches, Entrances and Exits are Stateless–Session–Beans. The Workers find their Benches by looking them up in the JNDI–Tree.

Of course it is possible to also implement other kinds of JavaBeans for the Workshop entities. E.g. it might be reasonable to implement a Bench as a Stateful–Session–Bean or even as an Entity–Bean. But up to now we did not face a situation where we need this.

We will see in the next Section how the KA is implemented upon the JWS. By this we will get a better understanding how the JWS framework works.

We have another implementation of the Workshop which is based on the JINI [12] technology. Currently we are about to implement a version based upon plain Java. These are out of the scope of this paper and are not further mentioned.

### D. The Key Authority

The KA was build upon the JWS framework. To explain the implementation of the KA we take a look at an exemplary procedure. We show step by step what happens in the KA when it generates an RSA [13] key pair, produces a respective X.509 certificate [14] and finally makes a PKCS#12 [15] token out of them. As the KA is offline, the communication is done by exchanging ITP–Messages [16] in files.

A special Entrance, called ITP–Entrance is scanning the respective directory for ITP files from the Registration Authority (RA). When a file is found the ITP–Entrance verifies its signature. If the signature is valid and from an authorized entity the ITP–Entrance reads and interprets the contained data. If the data is sound the ITP–Entrance instantiates an appropriate Worker and releases it to the system. Due to the kind of application in our case this is a PKCS12–Worker.

The PKCS12–Worker knows exactly what to do and which Benches it needs. The Worker holds a password for the PKCS12 token and a raw X.509–Certificate where the RA already filled in the user data. To generate the key pair the Worker must go to a RSA–Bench. It finds one by searching in the JNDI tree and goes there.

The RSA–Bench is a facade for a PCI crypto card which is able to generate secure RSA key pairs. The Worker requests a key pair from the RSA–Bench. This request is forwarded by the RSA–Bench to the underlying crypto hardware. The hardware generates the key pair and hands it over to the RSA–Bench. The Bench gives the key pair to the Worker. Having this, the Worker wanders to the Issuer–Bench.

The Issuer–Bench is a facade for the smart card containing the issuer private key. The PKCS12–Worker gives the raw certificate (now containing the public key) to the Issuer–Bench and requests a signature. The Bench checks the issuer name and completes the certificate with a valid serial number. Then it sends the certificate to the smart card for being signed. Having it back the Bench passes the signed certificate back to the PKCS12–Worker. Now the Worker has to get to the PKCS12–Bench.

The PKCS12–Bench offers services for creating PKCS#12 tokens. The Worker gives the key pair, the certificate and the password to the Bench. The Bench produces a PKCS#12 structure containing the key pair and the certificate and secures it with the given password. The created structure is given back to the Worker. Now the Worker has finished its job and searches an Exit.

Arrived at the ITP–Exit the Worker is destroyed and its data is extracted and packed into an ITP structure. The data consists of the created PKCS#12 structure and some meta data about the process. Finally the Exit sings the ITP structure and saves it as a file in the appropriate directory.

## IV. AUDIT

This Section shows how the various ilities are fulfilled by our design and its implementation.

### A. Explicit Ilities

*1) Availability:* The availability is realized by realizing the ilities robustness, performance and security.

*2) Configurability:* The application server provides a means to deal with configuration files. This is utilized. In addition it is possible to access those properties via the Java Management Extensions (JMX) [17]. We paid attention to use property variables instead of hard coded values at all reasonable places.

*3) Durability:* The flexibility in selecting cryptographic algorithms together with their providers is achieved by using the Java Cryptography Extension (JCE) [18]. This is the very purpose of the JCE framework.

*4) Failover:* The insecurity of ciphers (respective signatures) created with insecure keys or algorithms must be avoided. This can be achieved by applying multiple encryptions (respective signing) to the data. This is the topic of S. Maseberg's PhD thesis [19]. The KA supports this technology.

*5) Interoperability:* There are some default Workers which support common standards like X.509 certificates or revocation lists. Special wishes are implemented in dedicated Workers. The JWS concept makes it possible to conform to any standard or non standard formats.

*6) Manageability:* This is solved by utilizing the JMX technology. Therefor our code just has to accord to the M–Bean specification. As this is related to the EJB specification this is nearly for free. All the environmental tasks regarding the management are solved by the JMX framework.

*7) Modularity:* The application server is able to run in clustered mode. The server cares for transparency of distribution by hiding this fact from the enterprise beans. By deploying certain Workers and Benches on certain machines the partitioning is realized.

*8) Performance:* Again the application server solves this problem. High capacity systems are installed on clustered application servers with many instances of the respective Workers and Benches. Low capacity systems run on single machines.

*9) Reliability:* The JWS design enables an easy testing and an easy evaluation of the system. It splits the system in small functional parts. It is easy to test the whole system or only certain parts of it (e.g. testing of new Benches by implementing special Test–Workers). This can even be done with the productive installations at the customer's side. Additionally it is easy to evaluate the desired parts while ignoring the undesired ones. We experienced this while the system was evaluated to CC EAL 3 augmented.

*10) Retracability:* This is solved by implementing log messages. This is done utilizing the Log4J [20] technology. It is possible to configure various log levels and log formats. All the environmental tasks regarding the logging are solved by the Log4J framework.

*11) Robustness:* Software errors are handled by the individual Worker coping with the erroneous task. This is done by using the exception handling technology included in the Java language. Hardware errors are handled by the application server. In clustered mode the server is able to incorporate suitable failover mechanisms.

*12) Scalability:* This is also a task for the application server. While running in clustered mode it is possible to add and remove hardware dynamically. Workers and Benches can be deployed and undeployed dynamically, too.

*13) Security:* Access control and secure communications are realized using an internal PKI. Each operator and module the KA communicates with is provided with a key pair and a certificate. It is possible to let an electronic watchdog scan the executables and configuration files for unauthorized access. It is also possible to run the KA in offline mode. Further the operating system can be used to restrict access to the software. Clearly, the physical shielding has to be done by the physical environment. As the security requirements are different in each scenario we need the implicit ility adaptability to realize security.

## B. Implicit Ilities

*1) Adaptability:* This demand is met by JWS design. Setting up the desired functionalities means deploying respective Workers and Benches.

*2) Embedabilty:* This is solved by the JWS design. By deploying the respective Entrances and Exits the data import and export can be adapted to all desired formats. Suitable Workers and Benches enable the generation of suitable products. By partitioning the system it is possible to run the desired services at the desired locations.

*3) Extensibility:* This is also solved by the JWS design. Extending the system means implementing new types of Workers and Benches. The existing code can remain untouched.

*4) Maintainability:* As the JWS design is very intuitive it eases the maintenance of the code. It is easy for programmers to understand the structure of the code. As the code is split into many small blocks, it is easy to fix errors in buggy modules while leaving the correct blocks untouched.

*5) Platform independence:* This demand is solved by the Java programming language. Java programs are platform independent. Further there is a variety of Java APIs for connecting to various services or hardware in a platform independent manner. Examples for this are the JDBC architecture for database access or the PKCS#11 standard for accessing cryptographic tokens.

## V. PROOF OF CONCEPT

We present details from some real world projects where the KA is used. We mention some notably facts about the KA in the respective projects. Those examples show that the software is used in very different environments and for very different purposes. Thus, our concept proved to be proper and successful.

The KA was developed in an academic environment. We added or removed some experimental features sometimes. Casually we exchanged one technology with another one. Clearly the software grew over the time and the former versions are less complete than the later ones. Thus, the mentioned projects do not use the very same version of the KA.

## A. Project RegTP

In this project we had to develop the overall system for the new German national root CA. This included the design and the implementation of the workflow, the software, the hardware, the environmental issues and the organizational issues. The TC is hosted at the German Regulatory Authority for Telecommunications and Posts (RegTP). As this TC is the root for legally binding digital signatures in Germany it is a really high secure application. See [21] for the TCs homepage.

- The system uses two instances of the KA. One is for certification only (CertKA), the other one for revocation only (RevoKA). They operate absolutely independent from each other.
- Both have their respective signature key on chip card.
- For key generation a high evaluated third party key generator is used.
- It is a low capacity system. They issue about 10 certificates a year.
- The products conform to the ISIS–MTT standard respective the appendant SigG–Profile.
- The whole system runs in a strongroom which was build especially for that purpose.
- The system was evaluated to CC EAL 3 augmented. The strength of the established security mechanisms is "high". Details for this can be found in [22].
- The system is able to host multiple issuers. Thus, it is possible to additionally act as national root CA for foreign countries.

## B. Project JLU

The Justus Liebig University (JLU) of Giessen issues certificates to all students. Again we had to develop the overall

system, but we had some stringent demands on the workflow. This is a system with medium security settings. The respective homepage can be found at [23].

- The certification and the revocation are done with one KA using the same key.
- The KA is online and is hosted in the rooms of the computing center.
- The issuer key is on a chip card.
- The TC issues chip cards with pre–produced keys. The KA just certifies the public keys.
- It is a high capacity installation. They issue about 40.000 certificates a year. By using an automatic chip card personalization device it is possible to personalize 500 chip cards a day (8 hours).

*C. Project RBG*

The department of computer science of Technische Universität Darmstadt (RBG) issues certificates to all students and staff members of the department. We developed the overall system. It is a system with low security settings. This KA is based on the JINI version of the Workshop. The homepage of this trust center can be found in [24].

- Certification and revocation are done by the same KA with the same key.
- The KA is online and hosted in the rooms of the computing center.
- The issuer key is stored in a soft token.
- The key pairs are generated by the KA in software.
- The KA issues soft tokens.
- The KA runs in partitioned mode. The Entrance and the Exit run on the same machine as the RA. This machine is online. The Benches run on a separate machine with dedicated connection to the RA machine.
- This makes the KA running in a semi online mode.
- It is a medium capacity installation. They issue about 1500 certificates a year.

## VI. CONCLUSION

While designing the KA it arose that the J2EE framework lacks a metaphor for implementing applications in an intuitive and easy understandable way. In order to fill this gap we specified a suitable abstraction layer called Workshop. This specification is free from J2EE specific attributes and thus can also be used with other enterprise frameworks like .NET. Usage of the Workshop allows the programmer to utilize all services of the underlying system while even providing simplified access to ilities like Embedabilty or extensibility. The introduced metaphor is that of a real world Workshop. The Workers travel from Bench to Bench to fulfill their tasks. They bear the whole responsibility for the correctness of their work respective a for suitable error handling.

We implemented an instance of the Workshop upon the J2EE technology. The result is a framework called the JWorkshop. It is free of KA specific attributes. This enables the programming of arbitrary applications based on the JWorkshop. As a proof of concept we showed how we implemented the KA upon the JWorkshop. Using this example we demonstrated how the JWorkshop can be used to fulfill the KAs special demands. We saw that the Workshop metaphor at some points supports the realization of the mentioned explicit ilities. The major advantage of the Workshop is the support of the mentioned implicit ilities. Implementing enterprise applications upon the Workshop leads to an easy understandable code. Among other things this code is easily maintainable, easily adaptable and easily extensible.